# Design of a Peanut Hohlraum with Low Gas-Fill Density for the Laser Megajoule

**X. Li (李欣)[*], C. S. Wu (吴畅书) , Z. S. Dai (戴振生), D. G. Kang (康洞国)[†], W. D. Zheng (郑无敌), P. J. Gu (古培俊), P. Song (宋鹏)**

*Institute of Applied Physics and Computational Mathematics, Beijing 100094, China*

Recent experiments on the National Ignition Facility [D.E. Hinkel *et al.*, Phys. Rev. Lett. 117, 225002 (2016)] demonstrates that utilizing a long, large case-to-capsule ratio (=3) conventional cylindrical hohlraum at moderate gas-fill density (=0.6 mg/cm$^3$ $^4$He) improves the drive symmetry controaums has a little chance to achieve ignition at an acceptable energy level due to its small margin for the laser cone propagation. In this letter, a noncylindrical hohlraum, called as peanut hohlraum, using a larger case-to-capsule (=3.46) at lower gas-fill density (=0.3 mg/cm$^3$ $^4$He) is proposed to ignite a high-foot pusher capsule with a shorter pulse (~9ns). The peanut hohlraum requires about 2.5 MJ laser energy to achieve 306 eV peak drive temperature while the low-z plasma electron density on the inner cone path is maintained very low which results in little simulated Raman backscattering and the high-z bubble still stays away from the inner cone path without the laser absorption in it, which favor the drive symmetry control. Crossed-beam energy transfer is also neglectable because the crossing position is far away from the LEH. The peanut hohlraum can provide a good drive environment for capsule to achieve ignition, so it is undoubted that it will add to the diversity of ICF approaches.

The indirect-drive approach to ignition and high gain in ICF involves the use of hohlraums [1,2]. A hohlraum consists of a high Z case with laser entrance holes (LEHs). The laser beams entering the LEHs are efficiently converted into x rays at the beam spots on the hohlraum wall in order to create a uniform x-ray radiation bath. The high-Z cavity (hohlraum) is used to smooth the distribution of radiation seen by a fuel-filled capsule. A one-dimensional (1D) spherical implosion optimizes the fuel compression and burn. The cylindrical hohlraums are used most often in inertial fusion studies and are selected as the ignition hohlraums on NIF [3], which use CH capsules and shaped laser pulses that are 14-22 ns in duration. Typical capsule convergence ratios are 25 to 45, so that drive asymmetry can be no more than 1% [1], a demanding specification on hohlraum design. In the cylindrical hohlraums, the Legendre polynomial modes $P_2$ and $P_4$ of the flux on capsule are the main asymmetry modes required to be controlled. In theory, the asymmetry can be tuned to fulfill the specification of ignition capsule in the cylindrical hohlraums with two laser rings per side. However, the first ignition attempts used a 4-shock low-adiabat drive in cylindrical hohlraum at gas-fill density ($\rho$ = 0.96 mg/cm$^3$) and large outer-to-inner laser cone wavelength separation ($\Delta\lambda$) to control time-integrated x-ray drive asymmetry, leading to good compression [4] but lower neutron yield [5] than expected. The mix in hot spot due to the ablation front instability is the leading cause [6]. The recent work [7] on NIF observed fusion fuel gains exceed unity by using a 3-shock high-foot implosion method to reduce instability in the implosion in cylindrical hohlraums at higher gas-fill density ($\rho$=1.6 mg/cm$^3$) and still large outer-to-inner laser cone wavelength seperation ($\Delta\lambda$>5 Å) . However, the hot spot at band-time is still far from a sphere shown by the neutron image, and the analysis showed that the $P_2$ asymmetry of capsule is as high as -34%. The low-z gas tamps the motion of the wall, which can result in large x-ray drive symmetry swings in time, and helps to improve the symmetry of the implosion. However, high gas-fill density also leads to very complicated hohlraum environment. Significant plasma-induced laser backscatter (~40%-50%) on the inner cones [7], which is mainly composed of simulated Raman backscattering (SRS), has been observed on NIF, which make the drive asymmetry control (using beam phasing technology) useless. A crossed-beam energy transfer (CBET) [8] technique from the imposed large $\Delta\lambda$ is used to transfer a part of the laser energy of the outer cones to the inner cones in order to maintain the required time-integrated drive symmetry. However, CBET introduces spatiotemporal drive asymmetry. CBET results in spatially nonuniform laser beam spots generated by the volumetric overlap inner and outer cones at the LEH [9]. Further, simulation analyses show the amount of the energy transferred varies with time in the main pulse [9], contributing to the impairment of the inner cone propagation. Recently, NIF [10] utilizes a 2-shock 1 MJ pulse at the near-vacuum (NVH) platform ($\rho$ = 0.03 mg/cm$^3$) to drive a CH ablator capsule. A spherical capsule convergence with ignition-like high convergence ratio (=34) has been demonstrated by the high YOC ratio (~87%) and the very symmetric hot spot x-ray emission photograph at bang time. The fact, that there is little LPI observed in this experiment, shows that the combination of significant backscattering on the inner cones and large CBET mainly results in the drive asymmetry in cylindrical hohlraums.

Low gas-fill densities ($\rho \leqslant$ 0.6 mg/cm$^3$) are hopeful to improve the radiation drive environment. Laser backscatter levels along the inner cones should be reduced as well, because the cones are propagating through less dense plasma. Moreover, a lower gas-fill density should improve inner cone propagation due to less inverse bremsstrahlung absorption. Additionally, lower laser backscatter level will reduce the need for CBET. NIF experiments [11,12,13] have showed that using low gas-fill density is effective in controlling LPI. However, since lower gas-fill density results in an increased inward expansion of the hohlraum (wall motion of cool, dense plasma plus a hot under-dense "high z bubble"), increasing the hohlraum size is necessary to make the inward wall plasma stay out of the paths of the inner cones. A larger hohlraum (where the "case-to-capsule ratio", CCR $\equiv R_{hohlraum}/R_{capsule}$ increases) provides more room between the wall and capsule, allowing for improved inner beam propagation. Recent experimental results [13] with a longer, larger hohlraum (CCR=3) at lower gas-fill density of 0.6 mg/cm$^3$ (LLL) have showed significant improvements in the radiation environment, which result in an enhancement in implosion performance in driving a 3-shock high-adiabat capsule. The laser pulse lengths are in the range of 12-15 ns. Increasing the length of the hohlraum has been shown to improve P4 asymmetry [14]. Use of an LLL hohlraum mitigates SRS on the inner cones from 12.1% to 5.4%, reducing the reliance on CBET. The measured backscatter in LLL hohlraums is within NIF ignition requirements considering that the SBS level on the outer cones is 5.0% [14]. At the same time, in the LLL hohlraums, the level of hot electron preheat is 100×reduction compared with the level in the conventional cylindrical hohlraums. In a word, the experimental results show that the LLL hohlraums result in a neutron yield, fuel areal density and low-mode (P2) hot spot x-ray self-emission symmetry at peak compression with substantial improvement without CBET.

However, in the LLL hohlraums, averaged 16.2% of the inner cone energy is backscattered out of hohlraum due to SRS since the inner cone energy is about 1/3 of the total energy. It is easy for the SRS to rise to higher level because the level of 16.2% is very critical [15]. In the LLL hohlraum experiments [13], the peak radiation temperature is only 275 eV, which is much lower than the temperature required for the high-foot high-adiabat capsule [16] to ignite with high performance margin. Increasing the peak radiation temperature with more energy input can improve the 1D capsule performance directly. However, more dense low-z plasma occurring on the inner cones due to higher drive temperature is prone to simulate more SRS which will result in the lost of the drive symmetry control again. Additionally, simulation analysis [17] shows that the inner cones are being absorbed in the high-z bubble created by the outer cones for the LLL hohlraums. Use of lower gas fill density in order to reduce the new SRS risk can result in a larger high-z bubble, which increases the absorption of the inner cone energy in the bubble and makes the drive asymmetry tuning fail. Of course, increasing hohlraum sizes in equal proportion can provide more room for the inner cone propagation, which allow lower gas-fill density and larger high-z bubble. But, this method is inefficient due to more energy input required to maintain drive. And the input will exceed all the energy limits of the facilities in the world according to our estimations. The high-foot high-adiabat capsule [16] sacrifices 1D performance to the insensitivity to the 2D and 3D hydrodynamic defects in order to demonstrate performance close to 1D expectations. It is not a ignition capsule with enough margin since the instabilities in implosion have chance to appear again once the capsule is pushed to ignite. The technique of "adiabat shaping" [18,19] for ablation front instability control is also effective in performance improvement. However, this technique needs a longer pulse to reduce the fuel adiabat, which also results in more dense low-z plasma on the inner cones. In a word, there is little chance to ignite the 3-shock high-adiabat capsule [16] within the LLL hohlraums without the impairment of the inner cones at an acceptable energy level unless new techniques, which improve the target performance remarkably, are proposed.

1-D capsule performance benefits from longer pulse lengths with lower fuel adiabat, but hohllraum drive asymmetry control by reducing laser backscatter favor shorter pulses. To solve the problem, a high-adiabat capsule design [20], where a layer of high-density material is used as a pusher between the fuel and the ablator, has been proposed to improve the hot-spot pressure with a very short drive pulse. The capsule geometry and the drive temperature are summarized in Fig. 1. This capsule design uses a very high-adiabat 2-shock drive pulse (145eV foot temperature and 306eV peak temperature) to mitigate the ablative instability, while the high-density pusher (SiC) helps to increase the shell density at the maximum shell velocity, which results in the high hot-spot pressure. The ignition is achieved with 12.1 MJ yield. The primary risk of the capsule comes from the hydrodynamic instability of the pusher's outer surface and 2D simulations show that the instability is at a low level [20]. The remarkable advantage of the capsule design is that the drive pulse is very short (~9 ns), which is beneficial to the drive asymmetry control. Meanwhile, compared to the conventional high-foot high-adiabat capsule design [16], the pusher capsule can ignite with more margin due to the improvements in the hot-spot pressure and theoretical capsule yield.

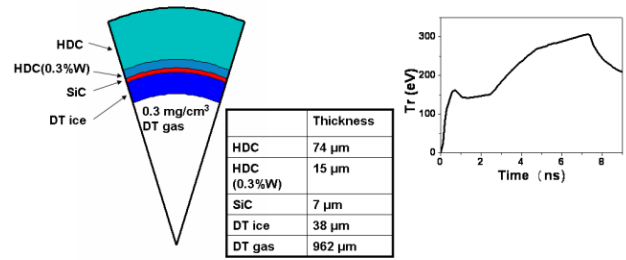

FIG. 1. The capsule geometry and drive temperature pulse.

In this Letter, a new hohlraum shape (see Fig. 2), called as peanut hohlraum, is proposed to drive the pusher capsule in order to achieve ignition at an acceptable energy level. Use of a peanut hohlraum over the conventional LLL cylindrical hohlraums gives a greater volume in the path of the inner cones, which allows for a larger high-z bubble or a lower gas-fill density. A peanut hohlraum designed for the pusher capsule is described in Fig. 2. The hohlraum material is pure Depleted Uranium (DU) [21], which can pride higher x-ray drive and less M-band x-ray than gold. The laser power is shaped to meet the radiation temperature requirement of implosion with a peak power of 530 TW and total laser energy of 2.5 MJ. The laser beam arrangement of the laser megajoule (LMJ) facility [22] is used. The full LMJ laser will be comprised of 240 beams grouped in 60 sets of 4 beams (quads) each. The 60 quads are arranged in three cones at angles 33.2° (inner cone), 49° (outer cone) and

59.5° (outer cone). Compared with the 23.5° inner cone in NIF [3], the 33.2° inner cone in LMJ can leave more room for high-z bubble growing when the inner cone beams are aiming at the waist-plane of hohlraum. And LMJ will provide more laser energy than NIF allowing for a longer and larger hohlraum. In this peanut hohlraum design, CCR is 3 at the hohlraum waist and 3.46 near the LEH, which leaves additional 500 μm large room for DU bubble created by the outer cone beams than the conventional LLL hohlraums which has only one CCR=3. It is the key geometry improvement which results that the peanut hohlraum has more margin for the inner cone propagation or the gas-fill density decreasing than that of the conventional LLL hohlraums. The hohlraum is filled with He gas at density 0.3 mg/cm$^3$, which is confined by a window over LEH of 0.5 μm thick polyimide. The capsule is irradiated by the x-emission rings created by the inner and outer cones on the interior wall respectively. In the two-ring-per-side illumination geometry, there are two zero-value positions of P$_4$ drive asymmetry contributed by laser ring, which are actually the nodes of the Legendre polynomial $P_4(\cos\theta)$ [23]. $\theta$ is defined as the angle under which the capsule "sees" the laser ring [24]. The ring closer to the waist plane of hohlraum is called the "inner ring", while the one closer to the LEH is called the "outer ring". Considering the plasma motion, the inner ring and the outer ring are initially placed at near the two P4 zero positions respectively to maintain the time-integrated P4 drive asymmetry small. As a result, the path of the inner cone are closer to the LEH edge than those of the outer cones in the peanut hohlraum. So the LEH has to be enlarged ($R_{LEH}$ = 2.465 mm, ~75% of the hohlraum waist radius) to prevent the absorption of the inner cone energy in the LEH edge.

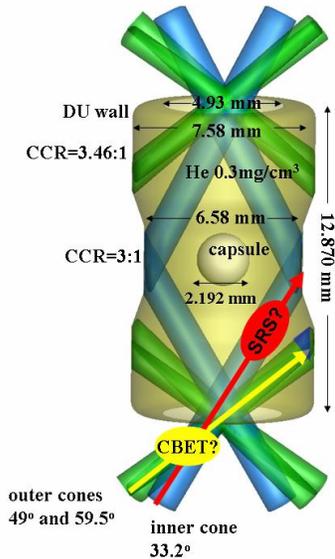

FIG. 2. The peanut hohlraum (left) for the pusher capsule is filled with He gas at low density (0.3 mg/cm$^3$), through which the inner cone propagate to the waist plane of hohlraum.

The energy balance [1] is used to relate the internal radiation drive temperature in hohlraum to the input laser energy by balancing the absorbed laser energy with the x-ray energy radiated into the wall, $E_W$, absorbed by the capsule, $E_C$, and the energy that escapes through the LEH, $E_{LEH}$, i.e.,

$$\eta_{CE}\eta_a E_L = E_W + E_C + E_{LEH} \qquad (1)$$

Where $\eta_a$ is the absorbed laser efficiency and $\eta_{CE}$ is the x-ray conversion efficiency from laser energy to soft x-rays. $E_W \propto \tau^{0.56}T_r^{3.45}A_W$, $E_C \propto \tau T_r^4 A_C$, and $E_{LEH} \propto \tau T_r^4 A_{LEH}$, where $\tau$ is the main pulse duration, $T_r$ the peak radiation temperature, $A_W$ the hohlraum wall area, $A_C$ the capsule area, and $A_{LEH}$ the LEH area. Usually, $\eta_{CE}$ is around 90% on NIF [25,26]. NIF experimental results [7] of the conventional cylindrical hohlraums show that $\eta_a$ is about 85%. The 15% loss is almost from the inner cone due to SRS during the main pulse, and the outer cones have little energy losses. In our target design, the SRS scatter on the inner cone is expected to be small (see below), so $\eta_a$~1. We apply this energy balance to the ignition peanut hohlraum and find that 2.3 MJ laser energy is required to produce 306 eV peak radiation temperature. The 2D non-equilibrium radiation hydrodynamics code LARED-Integration [28] is used to simulate the peanut hohlraum. Capsule needs a shaped radiation temperature to launch sequential shocks [1]. In our design, the laser power (Fig. 3) is tuned to meet the drive temperature requirement of the pusher capsule (Fig. 1). From simulations, the peak power and the total laser energy are 530 TW and 2.5 MJ, which is close to the energy estimated by energy balance. The peak M-band (>1.8 keV) fraction in the peanut hohlraum is about 18% and the M-band profile is also considered in the pusher capsule design procedure. During the design procedure of the peanut hohlraum by using 2D simulations, the time-dependent P$_2$ drive asymmetry is corrected by dynamically tuning the inner cone fraction (the ratio of the inner cone energy to the total delivered energy). An analytical model of the time-dependent P2 asymmetry for a two-ring-per-side illumination geometry[25] is used to provide the initial inner cone fraction. The best cone fraction at the peak power is about 28.5% which is lower than that (~33.3%) in the conventional cylindrical hohlraums [3] due to less inverse bremsstrahlung absorption resulted from the very low gas-fill density in the peanut hohlraum.

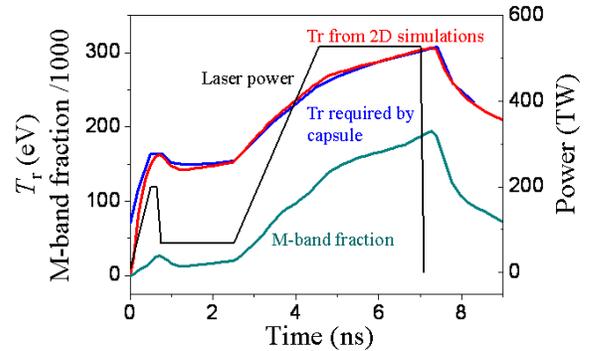

FIG.3. The hohlraum radiation temperature (red line) from 2D simulations, created by the laser pulse (black line), is almost identical to the drive temperature (blue line) required by the pusher capsule. The M-band fraction profile (green line) is drawn.

The greatest advantage of the peanut hohlraum is that the inner cone can propagate to the waist plane with little energy lost caused by SRS issue and high-z bubble absorption, which usually occur in the cylindrical hohlraums. Maps of the spatial distributions of the laser absorption, the electron temperature and the election density at the middle time of the main pulse are shown in Fig. 3. Although the gas-fill density is as low as 0.3 mg/cm$^3$, the DU bubble is still staying away from the inner cone path due to the high CCR (=3.46) close the LEH. The LEH is also large enough to allow the laser cones to propagate through without absorption in the LEH edge. According to experiments on NIF [7], the most critical risk for LPI in the peanut hohlraum is the SRS along the inner cone. SRS linear gain is defined as $G_{SRS} \propto \int I_L K_R(n_e/n_c, T_e) dl$, where $I_L$ is laser intensity, $K_R$ is integral kernel and strongly dependent on $n_e/n_c$ in the parameter space of concern. The integration is along the laser path. The SRS gain is widely used to estimate the SRS risk [1,29-31]. The intensity of the inner cone is ~$5\times10^{14}$ W/cm$^2$, which is similar to those used in the conventional cylindrical hohlraums [3]. SRS mainly occurs in the low-z gas due to small damping. Due to the motion of the wall, the gas is compressed by the wall and the ablator blow-off leading to a relative high-density (typically $n_e/n_c$~0.05 with very high gradient) and very small (~1 mm) region along the path of the inner cone. The high density gradient caused by capsule ablator coming into the path is very beneficial to the suppression of SRS [31]. The density on the other part of the inner cone path inside the low-z plasma drops quickly below 0.02. $T_e$ is ~2.5 keV along the most part of the inner cone path. To quantify the expected backscatter, we calculate the small-signal parametric gains along an ensemble of simulation rays using HLIP postprocessor [31]. HLPI uses the simulated electron temperature and density profiles to evaluate the LPI gain integrated along each ray path for a specified cone. We deduce an instantaneous reflectivity based on the Tang formula [32,33] and perform a convolution with the laser power history to extract a total backscatter reflectivity for the inner cone. For the inner cone, the SRS reflectivity is essentially zero which is consistent with the very low density along the path of the inner cone which results from the very low gas fill density initially. The HLIP analysis shows that SBS simulated mainly in the high-Z bubble is the main LPI risk on the outer cones in the peanut hohlraum. The laser intensity ($I_L$~$1\times10^{15}$ W/cm$^2$), temperature ($T_e$~3 keV), density ($n_e/n_c$<0.1) and bubble size ($L_p$~1 mm) are all similar to those of the conventional cylindrical hohlraums. Since experiments [7,13] on NIF do not show obvious SBS on the outer cones of the conventional cylindrical hohlraums, the SBS risk on the outer cones of the peanut hohlraum should be controlled. HLIP analysis also show that the SBS reflectivity of the outer cones is almost neglectable. Additionally, SBS can be mitigated by the introduction of boron to the wall liner, which increases the ion Landau damping. Experiments at the Omega laser by Paul Neumayer corroborate this mitigation technique [34]. That is why $\eta_a$ can be selected as ~1 before.

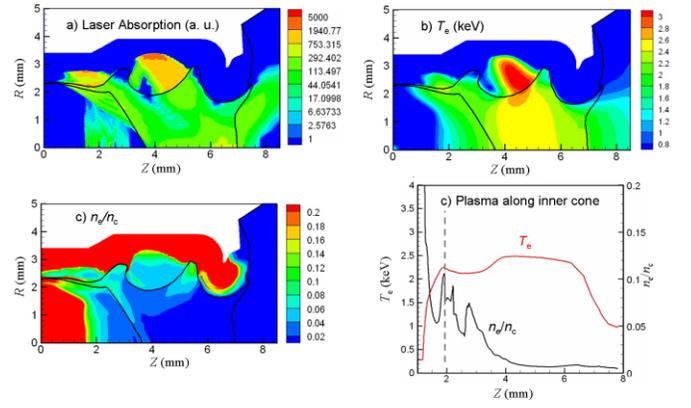

FIG. 3. Maps of plasma at the middle time of the main pulse: (a) Laser absorption (a. u.). (b) $T_e$ (in keV). (c) $n_e/n_c$. $n_c$ is the critical density. For a laser wavelength λ in microns, $n_c$ (cm$^{-3}$) = $1.1\times10^{21}/\lambda$ (μm)$^2$. (d) the plasma distribution along the inner cone path. The bold black lines in (a), (b) and (c) represent the boundaries of the wall, the capsule and the window. The grey dash line in (d) represents the boundary between the gass fill and the DU wall along the inner cone path.

In the conventional cylindrical hohlraums, the inner cones and the outer cones cross at the LEH. There is some CBET during the main pulse even at Δλ~0 due to the high plasma velocity at the crossing point. Suitable Δλ is usually required to suppress CBET. Then there is the potential for generating CBET during the picket pulse from the expanding LEH window even for low Δλ. However, in the peanut hohlraum, the crossing point is ~1.5 mm far away from the LEH. At this crossing point, there is no plasma during the picket pulse and the plasma density is almost zero during the main pulse. As a result, there is little CBET in the peanut hohlraum during the whole pulse.

In summary, a low gas-fill peanut hohlraum is proposed to allow for more room for the inner cone propagation without the impairment caused by SRS and high-z bubble absorption issues, which usually happens in the conventional cylindrical hohlraums. Although the LLL cylindrical hohlraums have achieved good hohlraum performance, the probability of ignition using the LLL cylindrical hohlraums on NIF is low unless the capsule and hohlraum performance can be greatly improved by new techniques. On the contrary, the peanut hohlraum allows very low gas-fill density to control LPI while the high-z bubble still stays away from the inner cone path avoiding possible inverse bremsstrahlung absorption. Using the peanut hohlraum to drive a pusher capsule, the ignition can be achieved with 2.5 MJ laser energy input at more performance margin. Based on the exploratory experiments on NIF, this target design uses some good parameters, including very low gas-fill density (=0.3 mg/cm$^3$), high CCR (=3.46), short pulse (~9 ns), high-adiabat et al., favoring both the drive asymmetry control (through improving the inner cone propagation) in hohlraum and the instability suppression (through the high-foot pulse) in capsule. The cost is more laser energy input, but the requirements, including the laser arrangement, can be fulfilled on LMJ.

This work is supported by the National Natural Science Foundation of China under Grants No. 11435011, Science Challenge Project under Grants No. JCKY2016212A505.

*  li_xin@iapcm.ac.cn
†  kang_dongguo@iapcm.ac.cn